\documentclass[journal=jpclcd,manuscript=letter,layout=twocolumn]{achemso}

\usepackage[version=3]{mhchem} 
\usepackage{amssymb,amsmath}
\usepackage{graphicx}
\usepackage[usenames,dvipsnames,svgnames,table]{xcolor}
\usepackage[normalem]{ulem}
\usepackage{hyperref}
\usepackage{siunitx}
\usepackage{bm}
\usepackage{textcomp}
\usepackage{gensymb}
\usepackage{placeins}
\usepackage{physics}
\usepackage{nicefrac}
\usepackage[caption=false]{subfig}
\usepackage{xcolor} 
\usepackage{cuted}

\newcommand{\kT}{k_{\mathrm{B}}T}
\newcommand{\fI}{$\mathrm{F}_{1}$}
\newcommand{\fo}{$\mathrm{F}_{\mathrm{o}}$}
\newcommand{\fofI}{$\mathrm{F}_{\mathrm{o}}\mathrm{F}_{1}$}
\newcommand{\new}{\color{red}}
\newcommand{\stkout}[1]{\ifmmode\text{\new \sout{\ensuremath{#1}}}\else{\new\sout{#1}}\fi}

\author{Emma Lathouwers}
\author{Joseph N.\ E.\ Lucero}
\author{David A.\ Sivak}
\email{dsivak@sfu.ca}
\affiliation{Department of Physics, Simon Fraser University, Burnaby, BC, V5A1S6 Canada}

\title{Nonequilibrium Energy Transduction in Stochastic Strongly Coupled Rotary Motors}


\begin{document}

\begin{tocentry}

\includegraphics[width=\textwidth]{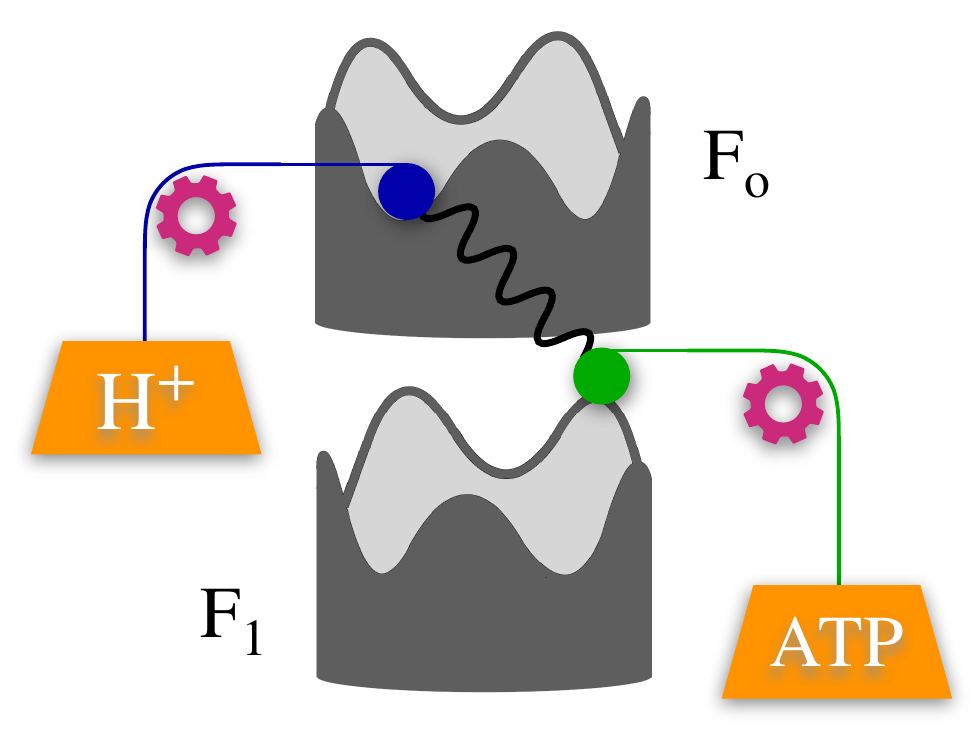}

\end{tocentry}

\begin{abstract}
Living systems at the molecular scale are composed of many constituents with strong and heterogeneous interactions, operating far from equilibrium, and subject to strong fluctuations. 
These conditions pose significant challenges to efficient, precise, and rapid free energy transduction, yet nature has evolved numerous molecular machines that do just this. 
Using a simple model of the ingenious rotary machine \fofI-ATP synthase, we investigate the interplay between nonequilibrium driving forces, thermal fluctuations, and interactions between strongly coupled subsystems.
This model reveals design principles for effective free energy transduction. 
Most notably, while tight coupling is intuitively appealing, we find that output power is maximized at intermediate-strength coupling, which permits lubrication by stochastic fluctuations with only minimal slippage.
\end{abstract}

Organisms must be continuously supplied with energy in order to persist~\cite{Schrodinger1944}.
Active research efforts focus on exactly how living things manage to effectively use input energy.
A major challenge stems from the fluctuating, soft matter nature of biological systems. 
Cells, organelles, and proteins all consist of many such components that are not rigidly coupled together. 
It is poorly understood how energy can be transduced effectively (rapidly and efficiently) in such a fluctuating, far-from-equilibrium setting.

Molecular motors play key roles in the harvesting, channeling, and consumption of energy in cells~\cite{Kolomeisky2007, Brown2017, Brown2019}. 
The input energy, often in the form of glucose, is in large part ultimately converted into the energy currency of living things, adenosine triphosphate (ATP); the subsequent dephosphorylation of energy-rich ATP molecules is an exergonic process that is coupled to cellular processes and thus drives a multitude of otherwise unfavorable reactions.

Most of the ATP produced during aerobic metabolism is made by the molecular motor \fofI-ATP synthase~\cite{Nelson2004}. 
The \fo~part of the motor harnesses a proton gradient across a membrane to rotate a central crankshaft, inducing a conformational change in the \fI~part, catalyzing the synthesis of ATP from adenosine diphosphate (ADP) and inorganic phosphate ($\mathrm{P}_\mathrm{i}$).
ATP synthase is believed to be an efficient energy transducer. 
In particular, the efficiency of \fI~(the ratio of the work done by the motor moving against a load to the energy stored in ATP) is estimated at $80-100\%$~\cite{Sumi2019,Toyabe2010}, and the efficiency of \fo\fI~is estimated at $65-90\%$~\cite{Silverstein2014}.
It is well established that this machine can rotate hundreds of times per second~\cite{Etzold1997,Ueno2005}, implying that the machine can deliver substantial power (producing hundreds of ATP per second). 

ATP synthase has been studied extensively, uncovering much about its structure and function~\cite{Boyer1997, Yoshida2001, Junge2015}.
Yet the mechanism by which it transduces energy is not fully understood. 
A common assumption in modeling \fofI-ATP synthase---and more generally in modeling molecular machines---is \emph{tight coupling}, where two coordinates are perfectly correlated, moving in lockstep.
In \fofI-ATP synthase this typically means tight mechanical coupling between the \fo~and \fI~subsystems~\cite{Ai2017}, tight mechanochemical coupling between \fo~and the proton current, and tight mechanochemical coupling between \fI\ and ATP synthesis and hydrolysis.
Tight coupling is not always mentioned explicitly, but instead a fixed stoichiometry between the number of protons translocated and each ATP synthesis/hydrolysis event is assumed~\cite{Anandakrishnan2016}.
There is often merit to these assumptions: 
Coupling that is too loose would be inconsistent with the bounds on efficiency, and tight coupling often reduces model complexity.
Soga et al.~\cite{Soga2017} find close agreement with tight coupling between \fo~and \fI~in thermophilic Bacillus PS3.
On the other hand, coupling that is merely strong (weaker than tight coupling, but still exhibiting significant correlation between coupled coordinates) is inevitable in many biological systems due to their fluctuating nature.
Slippage between the \fo~and \fI~subunits (producing proton current without concurrent ATP synthesis) has been observed~\cite{Toyabe2011,Feniouk1999}, hinting at intermediate-strength coupling between the proton current and ATP synthesis.

Recent years have seen growing interest in systems strongly coupled to the environment~\cite{Jarzynski2017,Seifert2016,Crooks2016} and subsystems strongly coupled together~\cite{Strasberg2017}.
Despite these theoretical efforts, relatively few models have been constructed that consider strong coupling in molecular motors:
Evstigneev et al.~\cite{Evstigneev2009} studied interacting overdamped Brownian particles in a tilted periodic potential;
Golubeva et al.~\cite{Golubeva2012} studied a class of 2D Brownian machines across a range of coupling strengths;
and Xie and Chen~\cite{Xie2018} proposed a kinesin model with strong mechanochemical coupling between stepping and ATP hydrolysis.
We are not aware of a systematic study of tunable coupling.

To elucidate how coupled stochastic systems can reliably transduce energy, we introduce a conceptually simple model of a stochastic strongly coupled rotary motor inspired by \fofI-ATP synthase and study its steady-state behavior. 
We find that the machine's output power is maximized at intermediate-strength coupling and at some (coupling-strength-dependent) phase offset between the components' energy landscapes.
Stronger coupling intuitively reduces slip, but (more subtly) requires stronger stochastic fluctuations to more simultaneously push both subsystems over energy barriers.
Both intermediate-strength coupling and a phase offset enhance output power by staggering (temporally and spatially, respectively) the activated transitions of the coupled machine components.

\begin{figure}[t]
    \centering
    \includegraphics[width=0.48\textwidth]{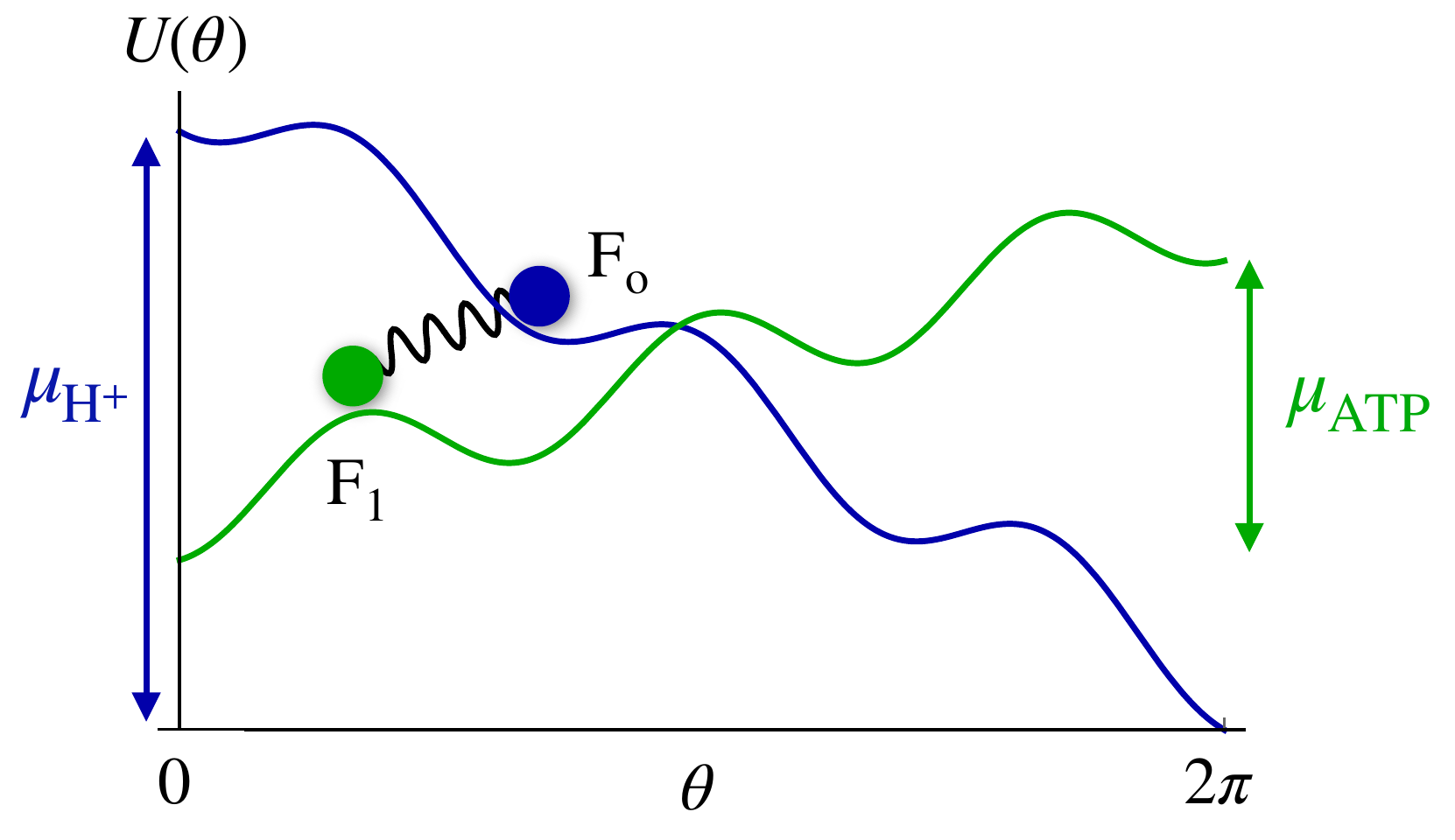}
    \caption{
    Effective energy landscapes experienced by each subsystem.
    Each subsystem diffuses on a periodic energy landscape and is coupled (depicted by a connecting spring) to the other subsystem.
    The tilt of each landscape reflects the effect of its respective chemical reservoir.
    }
    \label{fig:cartoonmodel}
\end{figure}

\fo~and \fI~are each modeled as a subsystem enacting a biased random walk on a periodic energy landscape, constrained by the other subsystem through an energetic coupling.
Figure~\ref{fig:cartoonmodel} shows a model schematic.
The two-dimensional energy landscape of the joint system is:
\begin{equation}
\label{eq:energylandscape}
\begin{aligned}
    V(\theta_{\rm o}, \theta_1) = 
    &-\tfrac{1}{2}E_{\rm o} \cos{ n_{\rm o} (\theta_\text{o} -\phi)} - \tfrac{1}{2}E_{1} \cos{ n_1 \theta_1 } \\
    &- \tfrac{1}{2}E_{\rm couple} \cos{ \left( \theta_\text{o} - \theta_1 \right) } \ .
\end{aligned}
\end{equation}
Here $\theta_{\rm{o},1}$ is the angular orientation of each subsystem, $\phi$ is the relative phase offset between the landscapes, $n_{\rm o,1}$ is the number of energy barriers of each energy landscape, $E_{\rm o,1}$ is the barrier height of each (untilted) landscape, and $E_{\rm couple}$ is the strength of the coupling favoring angular alignment of subsystems. 
\fo~and \fI~actually have many degrees of freedom, but to model inter-subsystem coupling, a sufficient statistic is the potential of mean force \cite{Frenkel2002} in the degrees of freedom (here $\theta_{\rm{o}}$ and $\theta_1$) that couple the subsystems.

The biased motion of each subsystem arises from (externally maintained) chemical driving forces.
The proton concentration difference driving ATP synthesis is modeled as a constant chemical driving force $\mu_\mathrm{H^+}$ pushing \fo~in one direction.
The free energy required to synthesize ATP by \fI~is modeled as a constant chemical driving force $\mu_\mathrm{ATP}$ pushing \fI~in the opposite direction.
Each full rotation of \fI~results in the synthesis of $n_1$ ATPs.
Such tilted energy landscapes with barriers constitute the dominant way of encapsulating in a simple model the experimental findings about \fo~\cite{Miller2013}, \fI~\cite{Kawaguchi2014}, and the complete ATP synthase~\cite{Xing2005}.
Just as \textit{in vivo} ATP synthase, this model can operate in reverse: Energy can be input to \fI~through ATP hydrolysis, driving \fo~to pump protons across a membrane.

The evolution of the joint probability distribution $P(\theta_{\rm o},\theta_1,t)$ over subsystem states $\theta_{\rm o}$ and $\theta_1$, subject to the described landscape and chemical driving forces, is governed by the Smoluchowski equation~\cite{VanKampen1992},
\begin{align}
	\pdv{}{t} P(\theta_{\rm o},\theta_1,t) = \dfrac{1}{\zeta}\Bigg[\pdv{}{\theta_{\rm o}} \left(\pdv{V}{\theta_{\rm o}} - \mu_{\rm H^+} \right) 
	+ \frac{1}{\beta} \pdv[2]{}{\theta_{\rm o}} \nonumber \\
	+ \pdv{}{\theta_1} \left(\pdv{V}{\theta_{1}} - \mu_{\rm ATP}\right) 
	+ \frac{1}{\beta} \pdv[2]{}{\theta_1} \Bigg] P(\theta_{\rm o},\theta_1, t) \label{eq:2dFPE} \ ,
\end{align}
where $\beta \equiv 1/(\kT)$, and friction coefficient $\zeta$ sets the timescale of the system dynamics. 
The joint system evolves under periodic boundary conditions ($\theta_{\rm o,1} \in [0, 2 \pi)$) until it reaches steady state, and we calculate the probability current, power, and efficiency from the steady-state probability distribution.
In the special cases of infinitely strong coupling, no coupling, or no energy barriers, the probability current reduces to exact expressions~\cite{Risken1996} (see SI: `Current under tight coupling' and `Current in the absence of energy barriers').

We assume tight coupling between the chemistry (proton gradient and ATP synthesis/hydrolysis) and mechanical motion of the system, and equal barrier heights $E_{\rm o}=E_1$ for \fo~and \fI.
Chemical driving forces are chosen such that the machine can harvest energy from the proton gradient and synthesize ATP: $\mu_{\rm H^+} > 0$, $\mu_{\rm ATP} < 0$, and $|\mu_{\rm H^+}| > |\mu_{\rm ATP}|$. 
Each subsystem has three energy barriers (and hence three metastable states or energy minima), $n_{\rm o} = n_1 = 3$.
In the limit of infinitely strong coupling between \fo~and \fI, a tightly coupled system (where \fo\ and \fI\ move in lockstep) is recovered, and the system is effectively one dimensional (see SI: `Current under tight coupling').
The probability currents are
\begin{align}
	\mathcal{J}_{\rm o} = \dfrac{1}{\zeta}\left[ \left(\mu_{\rm H^+} - \pdv{V}{\theta_{\rm o}}\right)P - \frac{1}{\beta} \pdv{P}{\theta_{\rm o}}\right] \\
	\mathcal{J}_1 = \dfrac{1}{\zeta}\left[ \left(\mu_{\rm ATP} - \pdv{V}{\theta_1}\right)P - \frac{1}{\beta} \pdv{P}{\theta_1}\right]
\end{align}
and the input/output powers are
$\mathcal{P}_{\rm H^+} = 2 \pi \mu_{\rm H^+} \langle \mathcal{J}_{\rm o} \rangle$ and $\mathcal{P}_{\rm ATP} = - 2 \pi \mu_{\rm ATP} \langle \mathcal{J}_1 \rangle$, where the angle brackets denote averages over both spatial degrees of freedom.

\begin{figure}[h!]
    \centering
    \includegraphics[width=0.45\textwidth]{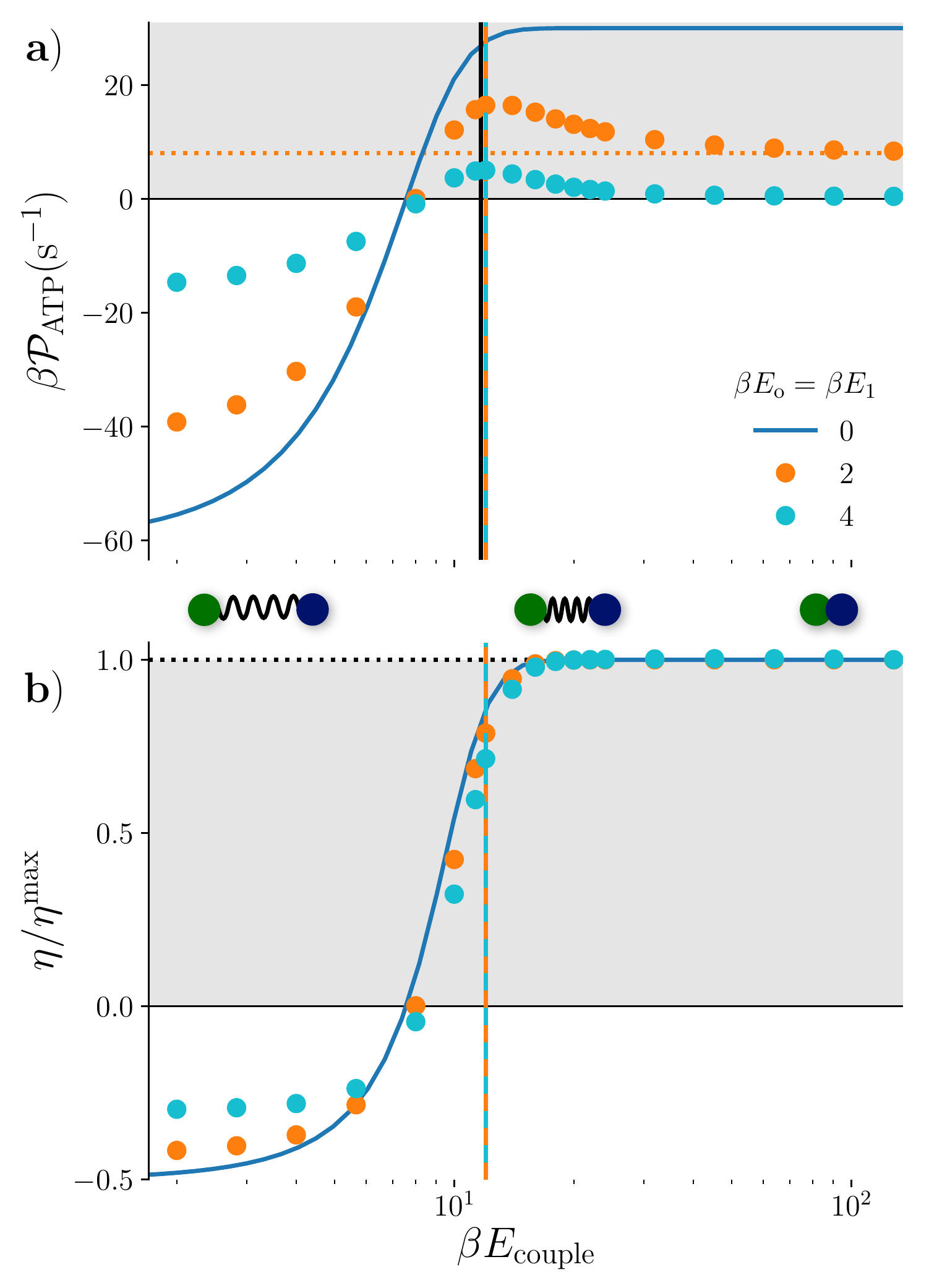}
    \caption{
    a) Output power as a function of coupling strength $\beta E_{\rm couple}$, without energy barriers ($\beta E_{\rm o}= \beta E_{1}=0$, blue solid curve) or with energy barriers ($\beta E_{\rm o}= \beta E_{1}=2$ [orange circles] or $4$ [light blue circles]).
    The phase offset is zero ($\phi = 0$), and chemical driving forces are $\mu_{\rm H^+} = 4\ \kT/ \rm rad$ and $\mu_{\rm ATP} = -2\ \kT/ \rm rad$.
    Horizontal dotted orange line: power output under infinite coupling for $\beta E_{\rm o}= \beta E_{1}=2$.
    Vertical dashed colored lines at $\beta E_{\rm couple} = 12$: coupling that approximately maximizes output power.
    Vertical black line at $\beta E_{\rm couple} = 11.7$: theoretical prediction~\eqref{eq:EcoupleMax} of coupling that maximizes output power.
    Grey shading indicates where the motor transduces energy.
    Schematics, left to right: weak, strong, and tight coupling.
    b) Efficiency under the same conditions, scaled by the theoretical maximum efficiency $\eta^{\rm max} = -\mu_{\rm ATP}/\mu_{H^+}$.
    Horizontal dotted line: maximum efficiency at infinite coupling.
    }
    \label{fig:flux_eff_Ecouple}
\end{figure}

Figure~\ref{fig:flux_eff_Ecouple} shows (a) the output power and (b) the efficiency, as a function of coupling strength, for systems with and without energy barriers. 
We expect the system to transduce energy (from the proton reservoir to the ATP reservoir) when the subsystems are coupled sufficiently strongly.
\fI~must move against its chemical driving force, which is achieved by coupling to \fo, itself pushed by its own chemical driving force.
When the coupling is too loose, the motion of each subsystem is predominantly determined by its own underlying landscape and chemical driving force, mostly unaffected by the other subsystem.
At sufficiently high coupling (here $\beta E_{\rm couple} \approx 8$), \fo\ drives \fI\ up its tilted landscape, resulting in energy transduction, see Fig.~\ref{fig:flux_eff_Ecouple}a.  
In the absence of energy barriers, the output power increases monotonically and plateaus at the infinite-coupling value (derivation in SI: `Current under tight coupling'),
\begin{align} \label{eq:power}
    \beta \mathcal{P}_{\rm ATP}^{\infty}\big|_{E_{\rm o,1}=0}
    = -\frac{\beta \mu_{\rm ATP} (\mu_{\rm H^+} + \mu_{\rm ATP})}{2 \zeta} \ .
\end{align}
In the presence of energy barriers, however, output power is maximized at intermediate-strength coupling, where slippage is minimal yet the subunits are sufficiently decoupled to allow temporal separation of their respective barrier crossings and hence greater current.  A simple theory accounting for the rates of the dominant events in transduction and slippage (SI: `Power-maximizing coupling' has details) 
predicts output power is maximized at coupling
\begin{align} \label{eq:EcoupleMax}
    \beta E_{\rm couple}^* &= \frac{4}{3} \ln 12 + \frac{4 \pi}{9} \beta \left( \mu_{\rm H^+} - \mu_{\rm ATP} \right) \ ,
\end{align}
agreeing well with the full numerical simulations.
In Fig.~\ref{fig:flux_eff_Ecouple}a, the maximum power greatly exceeds the power at infinite coupling, by a factor $\mathcal{P}_{\rm ATP}^{\rm max} / \mathcal{P}_{\rm ATP}^{\infty} \approx 13$ for 4-$\kT$~barriers, and $\mathcal{P}_{\rm ATP}^{\rm max} / \mathcal{P}_{\rm ATP}^{\infty} \approx 2$ for 2-$\kT$~barriers.

The efficiency in Fig.~\ref{fig:flux_eff_Ecouple}b is maximized for infinite coupling, both for systems with and without energy barriers.
Thus power and efficiency are not maximized simultaneously in a system with energy barriers, though the loss in efficiency can be small.

\begin{figure}[t]
    \centering
    \includegraphics[width=.48\textwidth]{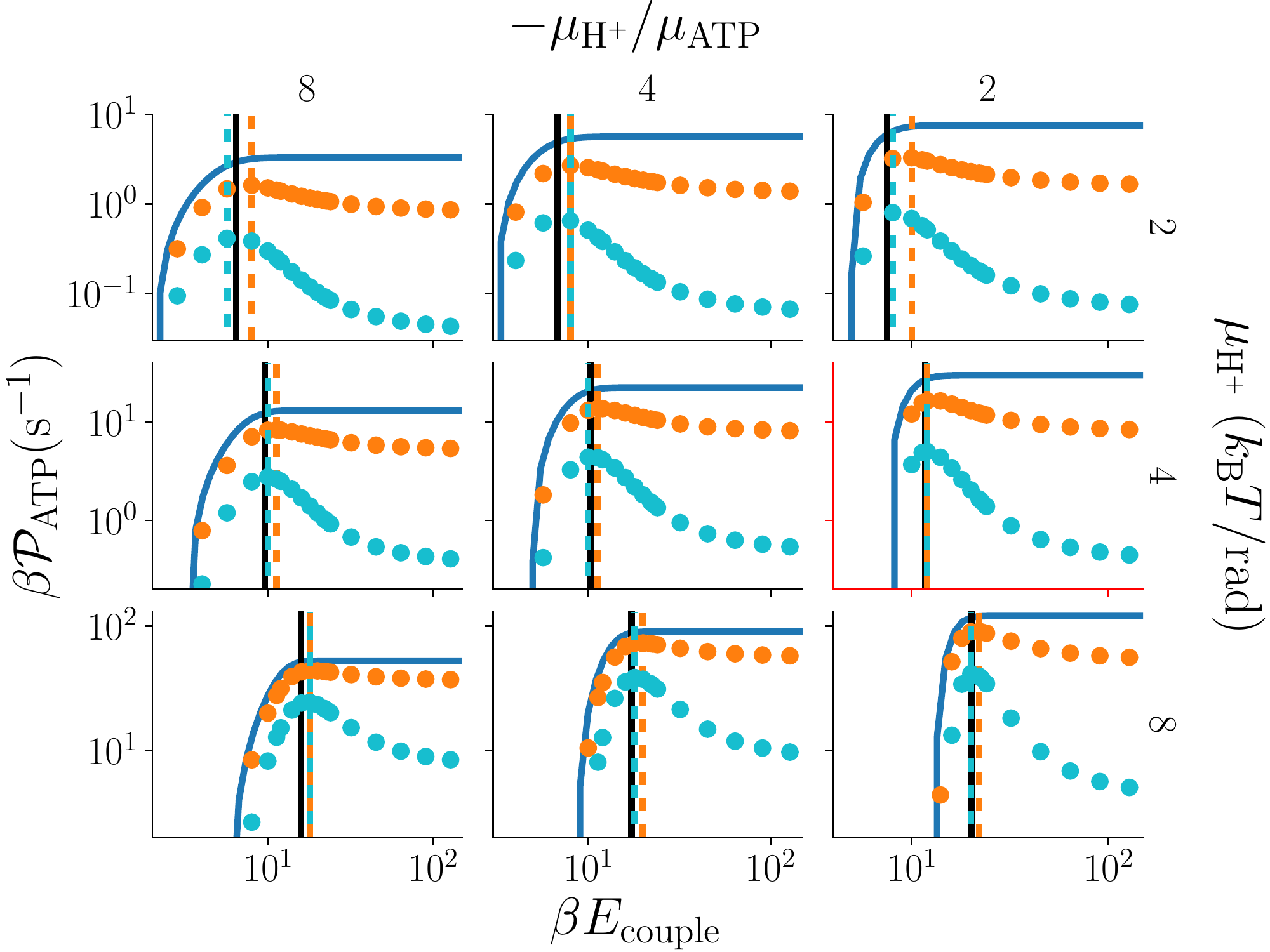}
    \caption{
    Output power as a function of coupling strength $\beta E_{\rm couple}$, without energy barriers ($\beta E_{\rm o}= \beta E_{1}=0$, blue solid curves), and with energy barriers ($\beta E_{\rm o}= \beta E_{1}=2$ [orange circles] or 4 [light blue circles]), for a variety of chemical driving forces.
    Vertical dashed lines: coupling strength that maximizes output power. Vertical black line: theoretical prediction of coupling strength that maximizes power output.
    Middle right sub-plot (red axes) reproduces Fig.~\ref{fig:flux_eff_Ecouple}a.
    }
    \label{fig:power_grid}
\end{figure}

Figure~\ref{fig:power_grid} shows the output power as a function of coupling strength for a variety of chemical driving forces.
It demonstrates that the result of output power being maximized at intermediate-strength coupling (and the accuracy of our simple theoretical prediction for power-maximizing coupling~\eqref{eq:EcoupleMax})
is robust to variation in chemical driving forces.

In the absence of energy barriers and at infinite coupling, the output power~\eqref{eq:power} increases monotonically with the proton driving force $\mu_{\rm H}^+$ (going down the rows in Fig.~\ref{fig:power_grid}), while the efficiency $\eta^{\rm max} = -\mu_{\rm ATP}/\mu_{H^+}$ decreases with increasing proton driving force $\mu_{\rm H}^+$.
The output power is maximized with respect to the ATP driving force $\mu_{\rm ATP}$ by $\mu_{\rm ATP} = -\frac{1}{2} \mu_{\rm H}^+$ (right column in Fig.~\ref{fig:power_grid}), which produces efficiency $\eta^{\rm max} = \frac{1}{2}$.
Consequently, output power is maximized in Fig.~\ref{fig:power_grid} (blue solid curve) in the lower right plot. 
The highest output power for systems with energy barriers (orange and blue circles) in Fig.~\ref{fig:power_grid} is also reached in the lower right plot. 
SI: `Barrier heights' explores output power variations with barrier height at rigid coupling.
The greatest gain in output power with respect to the infinite-coupling limit occurs in the upper right plot for 2-$\kT$~barriers, where $\mathcal{P}_{\rm ATP}^{\rm max} / \mathcal{P}_{\rm ATP}^{\infty} \approx 2$, and the middle right plot for 4-$\kT$~barriers, where $\mathcal{P}_{\rm ATP}^{\rm max} / \mathcal{P}_{\rm ATP}^{\infty} \approx 13$.

\begin{figure}[t]
    \centering
    \includegraphics[width=.48\textwidth]{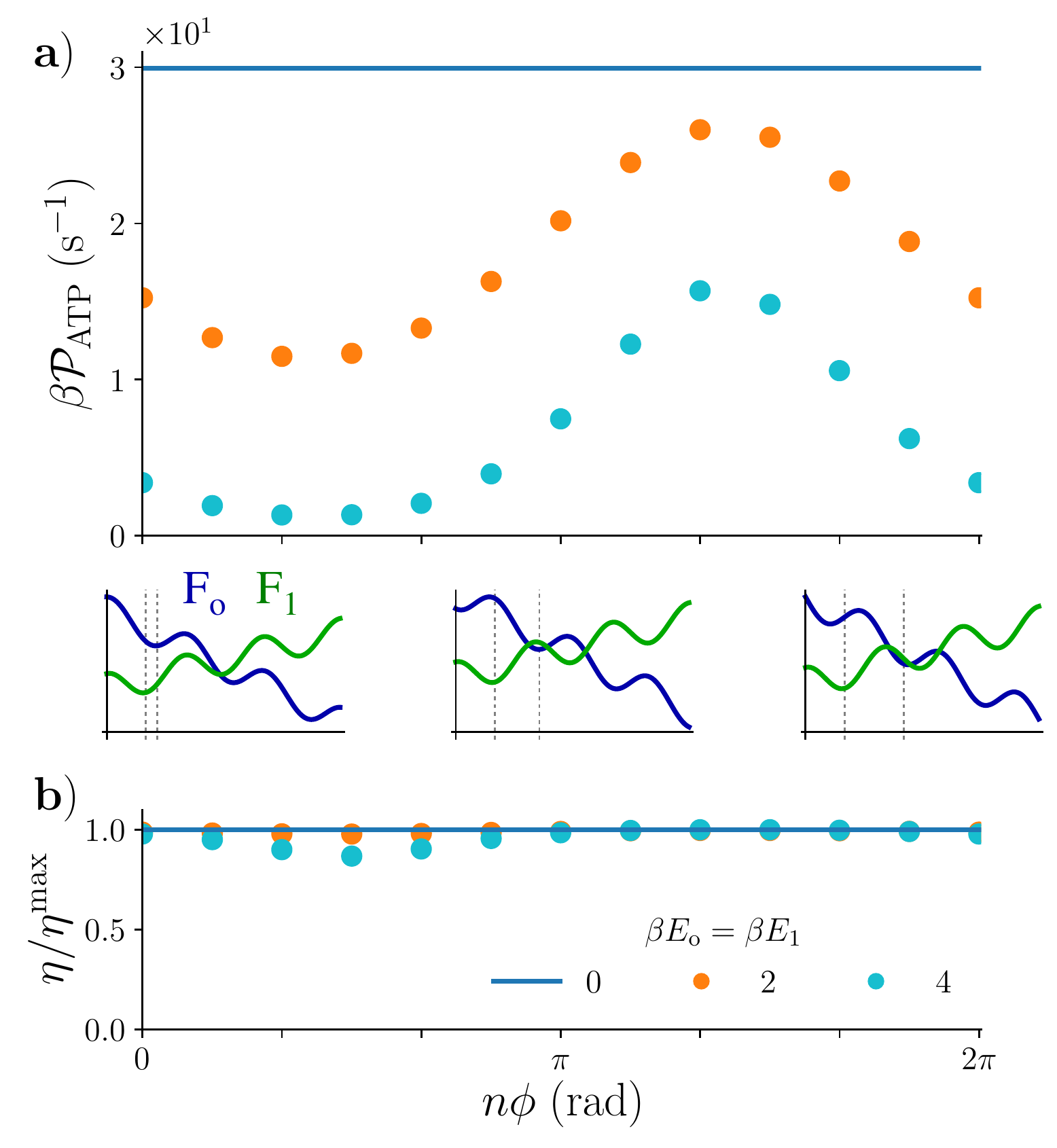}
    \caption{
    a) Output power as a function of scaled phase offset $n \phi$ at intermediate-strength coupling $\beta E_{\rm couple} = 16$, without energy barriers (blue solid curve) and with energy barriers ($\beta E_{\rm o}= \beta E_{1}=2$ [orange circles] or 4 [light blue circles]), for chemical driving forces $\mu_{\rm H^+} = 4\ \kT/ \rm rad$ and $\mu_{\rm ATP} = -2\ \kT/ \rm rad$.
    Schematics, left to right: Energy landscapes for small phase offset, medium phase offset (anti-aligned potentials), and large phase offset.
    Vertical dotted lines indicate phase offset.
    Tilt and barrier height are exaggerated for clarity. 
    b) Efficiency under the same conditions, scaled by the theoretical maximum efficiency $\eta^{\rm max} = -\mu_{\rm ATP}/\mu_{H^+}$.
    }
    \label{fig:power2}
\end{figure}

Figure~\ref{fig:power2} shows the (a) output power and (b) efficiency as a function of the (scaled) phase offset at strong coupling. 
Without energy barriers, the phase offset is physically immaterial, producing no variation.
With energy barriers and coupling strength $\beta E_{\rm couple} = 16$, output power is maximized at $n \phi \approx \nicefrac{4 \pi}{3}\ \mathrm{rad}$, qualitatively similar to the configuration illustrated in the rightmost schematic in Fig.~\ref{fig:power2}. Intermediate-strength coupling, such as $\beta E_{\rm couple} = 16$, permits sufficient flexibility that one subsystem can hop over an energy barrier while (temporarily) leaving behind the other subsystem.
A small phase offset makes it harder for a single subsystem to diffuse ahead to the next minimum, slowing the total system down and decreasing the output power.
At larger phase offset, staggering the energy barriers lowers the effective composite energy barrier, thereby increasing the output power.
Phase offset has minimal effect on efficiency (see Fig.~\ref{fig:power2}b).

\begin{figure}[t]
    \centering
    \includegraphics[width=.48\textwidth]{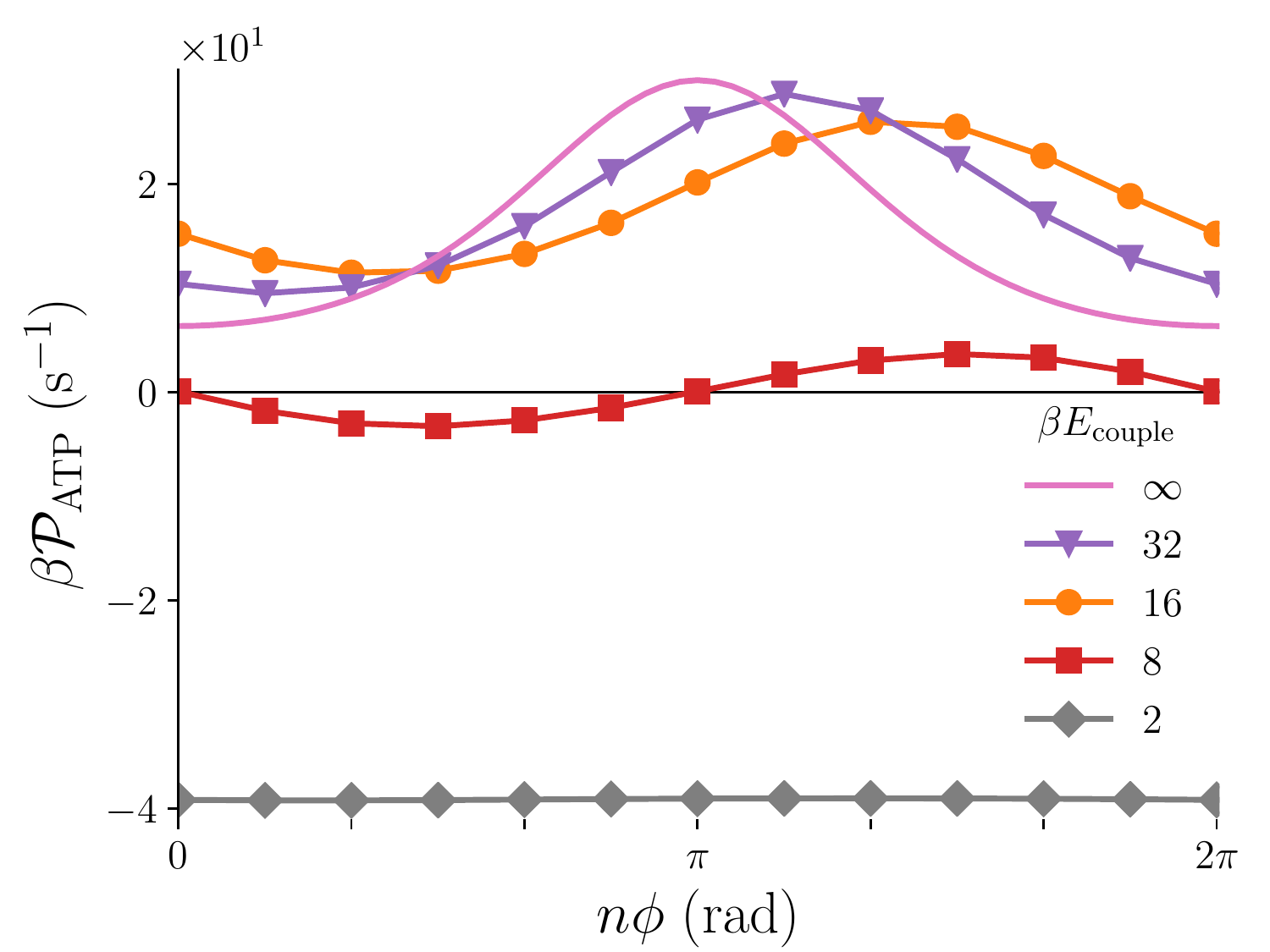}
    \caption{
    Output power as a function of the scaled phase offset $n\phi$ between subsystems, at coupling strengths $\beta E_{\rm couple}$ ranging from 2 to $\infty$, barrier heights $\beta E_{\rm o} = \beta E_1 = 2$, and chemical driving forces $\mu_{\rm H^+} = 4\kT/ \rm rad$ and $\mu_{\rm ATP} = -2\ \kT/ \rm rad$.
    An exact expression for the infinite-coupling result in Fig.~\ref{fig:power_phi} can be found in SI: `Current in the absence of energy barriers'.
    }
    \label{fig:power_phi}
\end{figure}

Figure~\ref{fig:power_phi} shows that the effect of varying the phase offset changes with coupling strength.
At very low coupling, the subsystems only weakly interact and are effectively independent, thus varying phase offset has no effect.
At low-to-strong coupling a relatively large offset is preferred.
At strong coupling, the phase offset that maximizes power output shifts towards $n \phi = \pi \ \mathrm{rad}$, corresponding to anti-aligned potentials (middle inset in Fig.~\ref{fig:power2}a).

These findings of output power being maximized at intermediate-strength coupling and at some coupling-strength-dependent phase offset are robust to varying the number of energy barriers (see SI: `Number of energy barriers') and should be preserved for even larger barrier heights (see SI: `Barrier heights').

Our results show that intermediate-strength coupling can lead to higher output power, while having only a small effect on efficiency (and producing scaled efficiencies comparable to experimental observations~\cite{Silverstein2014}).
This power-maximizing coupling strength reflects a trade-off between minimizing the slip between subsystems, and capitalizing on thermal fluctuations to kick subsystems over an energy barrier. 
Stronger coupling reduces slip, but reduces the lubrication by stochastic fluctuations, by more tightly tethering one subsystem to another and thereby reducing the propensity for a single 
subsystem to diffuse across an energy barrier.
Our results demonstrate that this behavior only arises in the presence of energy barriers (Fig.~\ref{fig:power_grid}), and when the subsystems have some mechanical freedom to explore different orientations.
In the absence of energy barriers the process is diffusion limited, whereas in the presence of energy barriers it is an activated process, where one has to wait for a sufficiently large thermal fluctuation to mobilize the system, which can be aided by looser coupling.

Due to tight coupling in the other transduction steps, our coupling strength represents the overall coupling of cross-membrane proton flow to ATP synthesis, so is upper bounded by the flexibility of any one of the components in \fofI's transduction chain. 
Experiments report that the $\gamma$-subunit (the primary mechanical coupling between \fo~and \fI) has an effective rotational spring constant 180 $\kT/\si{rad^2}$ in E.\ coli~\cite{Sielaff2008} and 54 $\kT/\si{rad^2}$ in Bacillus PS3~\cite{Okuno2010}.
These are 4-12$\times$ larger, and hence consistent as upper bounds, with our simple theoretical prediction~\eqref{eq:EcoupleMax} that at physiological conditions ($\mu_{\rm H^+} = 10~\kT/\rm rad$, $-\mu_{\rm H^+}/\mu_{\rm ATP} = 1.1$) the power-maximizing coupling is $30~\kT$, corresponding at small angular deviations to an effective spring constant $k_{\rm eff} = \tfrac{1}{2}E_{\rm couple} = 15~\kT / {\rm rad}^2$.

While Su\~ne and Imparato~\cite{Sune2019} studied the efficiency of a similar machine with no energy barriers, finding that tight coupling is necessary to approach reversible efficiency, our results imply that there is more to the story when energy barriers are 
present, which is in line with a number of stochastic machine models that consider strong coupling:
A maximum in probability current at strong coupling has been observed in a ratchet model for the mechanochemical coupling between molecular motors and ATP hydrolysis~\cite{Astumian1996}, in a Brownian heat engine composed of two coupled particles~\cite{Fogedby2017}, and in coupled stochastic oscillators~\cite{Zhao2004}. 
These findings suggest that intermediate-strength coupling can be valuable in a wide range of systems.
Here, we systematically study a range of coupling strengths between subsystems, and explicitly consider the consequences for the design principles of molecular machines.

A phase offset between the subsystem potentials can also enhance output power.
The optimal phase offset changes with coupling strength, because the coupling strength dictates the scale of inter-subsystem fluctuations.

Our results imply that output power is maximized by minimizing effective energy barriers.
This is analogous to the concept of splitting a large energy barrier into smaller barriers to improve speed~\cite{Wagoner2019}.
Both changes in the phase offset and coupling strength affect the effective energy barriers experienced by the composite system, the former spatially by changing the relative position of the barriers, and the latter temporally by allowing the subsystems to move asynchronously.

We focused on \fofI-ATP synthase to give context and concreteness to our model, however our findings are not limited to ATP synthase.
Any system of coupled stochastic subsystems with energy barriers can maximize power output by loosening the restriction of tight coupling and thereby reducing effective energy barriers.
More generally, we expect that the common assumption of tight coupling may overlook interesting phenomena in coupled stochastic systems. 
We confirm that minimal slip between subsystems is necessary for a highly efficient motor, as
ATP synthase is believed to be; however, we find that assuming perfectly tight coupling 
(which in any event is difficult to achieve in microscopic, fluctuating systems)
glosses over features that can contribute to driven biomolecular systems' functionality.

\begin{acknowledgement}
The authors thank Ralf Wittenberg (SFU Mathematics), John Bechhoefer, Jannik Ehrich, Steve Large, and Miranda Louwerse (SFU Physics) for insightful discussions and comments on the manuscript.
This work was supported by a Natural Sciences and Engineering Research Council of Canada (NSERC) Discovery Grant (D.A.S.), a Tier-II Canada Research Chair (D.A.S.), an NSERC Undergraduate Summer Research Award (J.N.E.L), and was enabled in part by support provided by WestGrid (\href{www.westgrid.ca}{www.westgrid.ca}) and Compute Canada Calcul Canada (\href{www.computecanada.ca}{www.computecanada.ca}).
\end{acknowledgement}

\begin{suppinfo}
Supporting information contains derivations of output power for infinitely strong coupling and for vanishingly small energy barriers, approximate theory for power-maximizing coupling,
computational methods, and results for varying numbers of energy barriers and varying barrier heights.
\end{suppinfo}

\clearpage
\appendix
\begin{strip}
	
\section{\huge Supporting Information}
\section{Current under tight coupling}
The Langevin equations describing the full system are
\begin{align}
	\zeta \dot{\theta_{\rm o}} 
	+ \frac{n_{\rm o}}{2} E_{\rm o} \sin n_{\rm o} (\theta_\text{o} -\phi) \label{eq:LangevinX}
	+ \frac{1}{2} E_{\rm couple} \sin{ \left( \theta_{\rm o} - \theta_1 \right) } 
	- \mu_{\rm H^+} &= \eta_{\rm \theta_{\rm o}},  \\
	\zeta \dot{\theta_1} 
	+ \frac{n_1}{2} E_1 \sin n_1 \theta_1 \label{eq:LangevinY}
	- \frac{1}{2} E_{\rm couple} \sin{ \left( \theta_{\rm o} - \theta_1 \right) }
	- \mu_{\rm ATP} &= \eta_{\rm \theta_1},
\end{align}
with the noise having statistics that satisfy
\begin{align}
	\langle \eta_{\rm i}(t) \eta_{\rm j}(t') \rangle = 2 \zeta \kT \delta_{\rm ij} \delta(t-t').
\end{align}
Here, over-dots denote a time derivative.
We impose identical friction coefficients $\zeta$.

We consider the case of equal numbers of energy barriers in each subsystem, $n_{\rm o} = n_1 = n$.
Summing \eqref{eq:LangevinX} and \eqref{eq:LangevinY}, dividing by two, and substituting $\theta_{\rm o} = \theta_1 = \theta$ (in the infinite-coupling limit of $E_{\rm couple} \to \infty$, the subsystems maintain the same angle), gives
\begin{align} \label{eq:rigid}
	\zeta \dot{\theta} 
	+ \frac{n}{4} E_{\rm o} \sin n (\theta -\phi) 
	+ \frac{n}{4} E_1 \sin n \theta
	- \frac{1}{2} (\mu_{\rm H^+} + \mu_{\rm ATP}) = 
	\eta_{\rm \theta},
\end{align}
where the composite noise term $\eta_{\rm \theta}$ satisfies
\begin{align} \label{eq:noise}
	\langle \eta_{\rm \theta}(t) \eta_{\rm \theta}(t') \rangle = \zeta \kT \delta(t-t').
\end{align}

The resulting one-dimensional energy landscape is
\begin{subequations}
	\begin{align} 
		V(\theta) &=
		-\frac{1}{4} E_{\rm o}\cos n( \theta - \phi) 
		- \frac{1}{4} E_{1} \cos n \theta \\
		&= -\frac{1}{4} E \cos n ( \theta - \varphi) \label{eq:landscape} \ ,
	\end{align}
\end{subequations}
for
\begin{align}
	E &\equiv \sqrt{ E_{\rm o}^2 + E_{1}^2 + 2 E_{\rm o} E_{1} \cos n \phi } \ , \\
	\tan n\varphi &\equiv \frac{ \sin n\phi } {\frac{E_{\rm o}}{E_{1}} + \cos n\phi} \ .
\end{align}
The chemical driving force exerted on the one-dimensional system is $\mu \equiv \frac{1}{2} (\mu_{\rm H^+} + \mu_{\rm ATP})$.
The exact expression for the average probability current is~\cite{Risken1996}:
\begin{align} \label{eq:fluxrisken}
	\langle \mathcal{J} \rangle = 
	\frac{\kT}{\zeta} \left[ (1-e^{-2\pi \beta \mu})^{-1} \int_0^{2\pi} \mathrm{d}\theta \, e^{\beta U(\theta)} \int_0^{2\pi} \mathrm{d}\theta' \, e^{-\beta U(\theta')} 
	- \int_0^{2\pi} \mathrm{d}\theta \, e^{-\beta U(\theta)} \int_0^{\theta} \mathrm{d}\theta' \, e^{\beta U(\theta')} \right]^{-1} \ ,
\end{align}
We integrate this numerically, using Mathematica's NIntegrate function with the ``DoubleExponential'' method. 
$U(\theta) \equiv V(\theta) - \mu \theta$ is the combination of the underlying energy landscape and the chemical driving force.
We note that when the number of energy barriers is not equal ($n_{\rm o} \neq n_1$), this result can still be used; the only difference lies in the energy landscape $V(\theta)$, which generally does not simplify as in \eqref{eq:landscape}.

In the limit of small energy barriers compared to the chemical driving force ($n_{\rm o} E_{\rm o}, n_1 E_1 \ll \tfrac{1}{2} (\mu_{\rm H^+} + \mu_{\rm ATP})$), also referred to as the case of no energy barriers, \eqref{eq:rigid} reduces to
\begin{align} \label{eq:infzero}
	\zeta \dot{\theta} 
	- \tfrac{1}{2} (\mu_{\rm H^+} + \mu_{\rm ATP}) = 
	\eta_{\rm \theta} \ ,
\end{align}
which describes diffusion subject to a constant force $\tfrac{1}{2} (\mu_{\rm H^+} + \mu_{\rm ATP})$.
Integrating \eqref{eq:infzero} over fluctuations gives the average drift velocity
\begin{align}
	\langle \dot{\theta} \rangle = \frac{\mu_{\rm H^+} + \mu_{\rm ATP}}{2 \zeta}.
\end{align}
Finally, output power is $\mathcal{P}_{\rm ATP} = - \mu_{\rm ATP} \langle \dot{\theta} \rangle = - 2 \pi \mu_{\rm ATP} \langle \mathcal{J} \rangle$.

\section{Current in the absence of energy barriers}
Starting from the dynamical equations for the full two-dimensional system~(\ref{eq:LangevinX},\ref{eq:LangevinY}), consider the limit of energy barriers much smaller than the chemical driving forces ($n_{\rm o} E_{\rm o} \ll \mu_{\rm H^+}$, $n_1 E_1 \ll \mu_{\rm ATP}$), leading to
\begin{align}
	\zeta \dot{\theta}_{\rm o} + \frac{1}{2} E_{\rm couple} \sin{ \left( \theta_{\rm o} - \theta_1 \right) } - \mu_{\rm H^+} &= \eta_{\rm \theta_{\rm o}} \ , \label{eq:LangO} \\
	\zeta \dot{\theta}_1 - \frac{1}{2} E_{\rm couple} \sin{ \left( \theta_{\rm o} - \theta_1 \right) } - \mu_{\rm ATP} &= \eta_{\rm \theta_1} \ . \label{eq:Lang1}
\end{align}
Summing and subtracting these equations, and changing variables to the mean
\begin{align}
	\bar{\theta} \equiv \frac{1}{2} ( \theta_{\rm o}+\theta_1 ) \ ,
\end{align}
and relative angle
\begin{align}
	\Delta \theta &\equiv \frac{1}{2} ( \theta_{\rm o}-\theta_1 ) \ ,
\end{align}
uncouples (\ref{eq:LangO},\ref{eq:Lang1}) to give independent Langevin equations:
\begin{align}
	\zeta \partial_t \bar{\theta} 
	- \frac{1}{2}( \mu_{\rm H^+} + \mu_{\rm ATP}) &= \eta_{\bar{\theta}} \ , \label{eq:CM} \\
	\zeta \partial_t \Delta \theta 
	+ \frac{1}{2} E_{\rm couple} \sin 2 \Delta \theta 
	- \frac{1}{2} ( \mu_{\rm H^+} - \mu_{\rm ATP}) 
	&= \eta_{\Delta \theta} \ . \label{eq:diff}
\end{align}
The transformed noise terms $\eta_{\bar{\theta}} \equiv \tfrac{1}{2}(\eta_{\theta_{\rm o}} + \eta_{\theta_1})$ and $\eta_{\Delta \theta} \equiv \tfrac{1}{2}(\eta_{\theta_{\rm o}} - \eta_{\theta_1})$ each satisfy
\begin{align} \label{eq:noise2}
	\left\langle \eta(t)\eta(t') \right\rangle 
	= \zeta \kT \delta(t-t') \ .
\end{align}

\eqref{eq:CM} can be integrated directly, exactly like \eqref{eq:infzero}.
\eqref{eq:diff} is analogous to \eqref{eq:rigid} in the sense that both are Langevin equations describing a system subject to a periodic energy landscape and a constant driving force. 
An exact expression for the probability current can be derived using \eqref{eq:fluxrisken}, where $U(\theta) = - \frac{1}{4} E_{\rm couple} \cos 2 \theta - \mu \theta$ and $\mu = \frac{1}{2} ( \mu_{\rm H^+} - \mu_{\rm ATP})$. 
The system is now reduced to one dimension with energy landscape $U(\theta)$ and an externally set chemical driving force $\mu$.
Linearly combining these probability currents gives each subsystem's probability current:
\begin{align}
	\langle \mathcal{J}_{\rm o} \rangle &= \langle \mathcal{J}_{\bar{\theta}} \rangle + \langle \mathcal{J}_{\Delta \theta} \rangle \ , \\ 
	\langle \mathcal{J}_1 \rangle &= \langle \mathcal{J}_{\bar{\theta}} \rangle - \langle \mathcal{J}_{\Delta \theta} \rangle \ .
\end{align}
Output power is $\mathcal{P}_{\rm ATP} = -\mu_{\rm ATP} \langle \dot{\theta}_1 \rangle = - 2 \pi \mu_{\rm ATP} \langle \mathcal{J}_1 \rangle$.

\end{strip}

\begin{figure*}[ht]
	\centering
	\includegraphics[width=\textwidth]{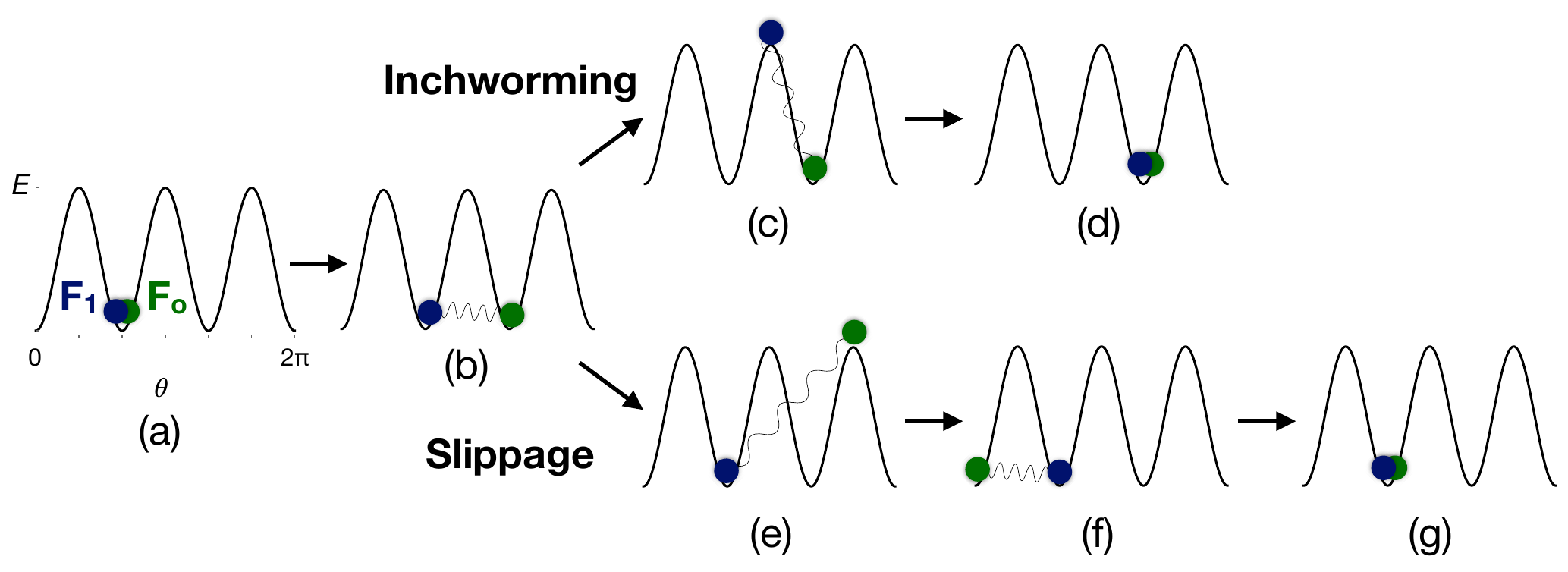}
	\caption{Visualization of inchworming and slippage.
		\fo\ and \fI\ start in the same state (a), then \fo\ advances to the next state (b).
		Inchworming, which leads to energy transduction, involves \fI\ advancing to catch up with \fo\ (c-d).
		Slippage occurs when \fo\ further advances to the next state (e-f), after which it is likely to continue on to complete a full cycle (g). 
		Throughout, $n=3$ and there is no phase offset. 
		Landscape tilts are omitted to simplify depiction. 
	}
	\label{fig:inchworm_slip}
\end{figure*}

\begin{figure*}[ht]
	\centering
	\includegraphics[width=0.5\textwidth]{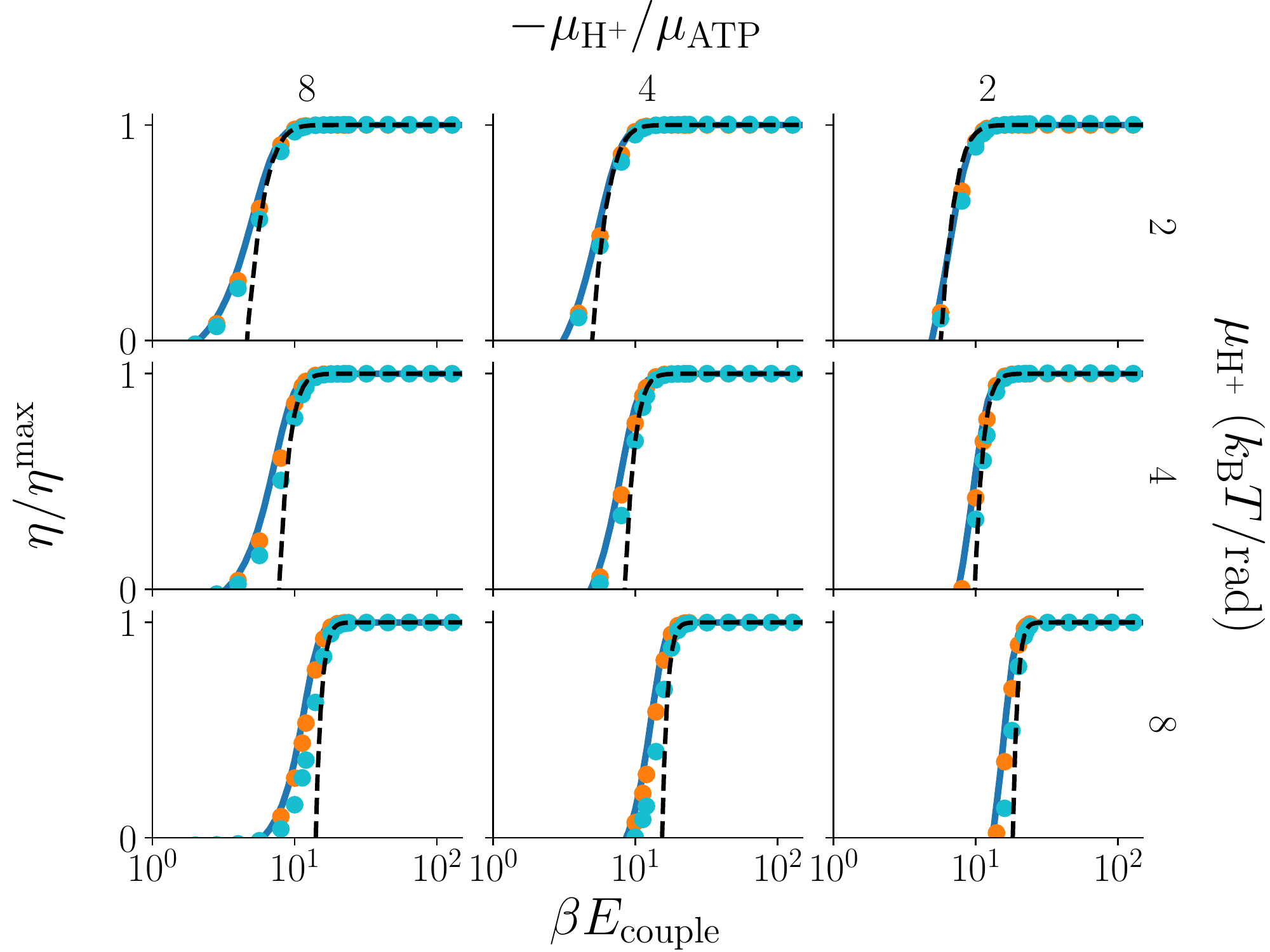}
	\caption{\label{fig:eff}
	Scaled efficiency as a function of coupling strength, for no barriers (blue curve), simulated for barrier heights $\beta E_{\rm o}= \beta E_{1}=2$ (orange circles) or 4 (light blue circles), and for simple theory (Eq.~\eqref{eq:effEcouple}, dashed black curve).}
\end{figure*}

\begin{strip}
\section{Power-maximizing coupling}

Here we present a simple theory to predict the coupling at which output power is maximized, by approximating its two components (input power and efficiency) in terms of the rates of the rate-limiting steps for energy transduction and slippage.  We restrict our attention to $n=3$ metastable states, no phase offset, and the regime of biological interest where the proton driving force is greater than and opposite in sign to the ATP driving force. 
	
For significant coupling, \fo\ and \fI\ are frequently in the same state (Fig.~\ref{fig:inchworm_slip}a). 
For tight coupling, \fo\ and \fI\ must cross a barrier simultaneously; but for somewhat weaker coupling, the most likely event from this configuration is \fo\ moving ahead to the next state (Fig.~\ref{fig:inchworm_slip}b).
From there, the two most likely events are `inchworming' where \fI\ catches up to \fo\ (Fig.~\ref{fig:inchworm_slip}c-d) or `slippage' where \fo\ moves further ahead (Fig.~\ref{fig:inchworm_slip}e-f), from which \fo\ will most likely continue further to its original position (Fig.~\ref{fig:inchworm_slip}g-h) without \fI\ having moved.
	
The rates $r=r_0 e^{-\beta \Delta E}$ of these competing steps are proportional to the exponentials of the respective energy barriers (Fig.~\ref{fig:inchworm_slip}c,e).
The rate-limiting step for inchworming (Fig.~\ref{fig:inchworm_slip}b$\to$c) has rate
\begin{align} \label{eq:inch1}
	r_{\rm inch} = r_0 \exp \left\{ -\beta \left[ E_1 - \frac{1}{2} E_{\rm couple} - \frac{\pi}{3} \mu_{\rm ATP} \right] \right\} \ .
\end{align}
The inchworming rate increases with decreasing barrier height, increasing coupling strength, or increasing ATP driving force.
The rate-limiting step for slippage (Fig.~\ref{fig:inchworm_slip}b$\to$e) has rate
\begin{align} \label{eq:slip1}
	r_{\rm slip} = r_0 \exp \left\{ -\beta \left[ E_{\rm o} + \frac{1}{4} E_{\rm couple} - \frac{\pi}{3} \mu_{\rm H^+} \right] \right\} \ .
\end{align}
The slippage rate increases with decreasing barrier height, decreasing coupling strength, or increasing proton driving force.
	
The ratio of inchworming and slippage rates is
\begin{align}
	\frac{r_{\rm slip}}{r_{\rm inch}} = \exp \left\{ \beta \left[ E_1  - E_{\rm o} + \frac{\pi}{3} \left( \mu_{\rm H^+} - \mu_{\rm ATP} \right) - \frac{3}{4} E_{\rm couple} \right] \right\} \ .
\end{align}
For identical \fo\ and \fI\ barrier heights, this reduces to
\begin{align}
	\label{eq:rRatio}
	\frac{r_{\rm slip}}{r_{\rm inch}} = \exp \left\{ \beta \left[ \frac{\pi}{3} \left( \mu_{\rm H^+} - \mu_{\rm ATP} \right) - \frac{3}{4} E_{\rm couple} \right] \right\} \ .
\end{align}
The slippage-inchworming ratio increases with decreasing coupling strength and with increasing magnitude of driving forces (i.e., more positive $\mu_{\rm H^+}$ or more negative $\mu_{\rm ATP}$).
Notice that barrier height has an identical effect on both rates, so cancels out when comparing the two. 
	
One inchworm event rotates the joint system 1/3 of a full cycle, whereas one slippage event results in a full cycle of slip, so the efficiency is simply expressed in terms of the rate ratio,
\begin{align}
	\frac{\eta}{\eta_{\rm max}} &= \frac{\frac{1}{3}r_{\rm inch} - r_{\rm slip}}{\frac{1}{3}r_{\rm inch}} \\
	&= 1 - 3\frac{r_{\rm slip}}{r_{\rm inch}} \ .
\end{align}
Substituting \eqref{eq:rRatio} gives the efficiency as a function of the coupling strength,
\begin{align}
	\label{eq:effEcouple}
	\frac{\eta}{\eta^{\rm max}} = 1 - 3 \exp \left\{ \beta \left[ \frac{\pi}{3} \left( \mu_{\rm H^+} - \mu_{\rm ATP} \right) - \frac{3}{4} E_{\rm couple} \right] \right\} \ .
\end{align}
Figure~\ref{fig:eff} shows that this simple theory accurately predicts the coupling at which efficiency begins to drop, across the examined variation of driving forces.  

\end{strip}
	
\begin{strip}
	
	For strong proton driving force, \fo\ backsteps are negligible, and for intermediate-or-stronger coupling, the slowest step is \fo\ stepping one state ahead of \fI\ (Fig.~\ref{fig:inchworm_slip}a$\to$b). Hence the (net) input power is simply proportional to the exponential of the activation energy of this slowest step,
	\begin{align}
		\mathcal{P}_{\rm H^+} \propto \exp \left\{ - \beta \left[ \frac{1}{4} E_{\rm couple} + E_{\rm o} - \frac{\pi}{3} \mu_{\rm H^+} \right] \right\} \ .
	\end{align}
	Thus the output power is 
	\begin{align}
		\mathcal{P}_{\rm ATP} &= -\frac{\eta}{\eta^{\rm max}} \mathcal{P}_{\rm H^+} \\
		&\propto \exp \left\{ - \frac{1}{4} \beta E_{\rm couple} \right\} - 3 \exp \left\{\beta \left[ \frac{\pi}{3} \left( \mu_{\rm H^+} - \mu_{\rm ATP}  \right) - E_{\rm couple} \right] \right\} \ .
	\end{align} 
	This predicts that output power is maximized at coupling strength
	\begin{align}
		\beta E_{\rm couple}^* &= \frac{4}{3} \ln 12 + \frac{4 \pi}{9} \beta \left( \mu_{\rm H^+} - \mu_{\rm ATP} \right)  \ .
	\end{align}
	
	This aligns with intuition: increasing either the driving force on \fo\ (more positive $\mu_{\rm H^+}$) or the resistive force on \fI\ (more negative $\mu_{\rm ATP}$) increases the slip between subsystems, hence the optimal coupling (that maximizes inter-subunit flexibility subject to only minimal slippage) shifts higher. 
	Figure 3 in the main text 
	shows that this prediction closely approximates the power-maximizing coupling in full numerical simulations. 
\end{strip}

\section{Computational methods}
All numerical code is freely available at Github~\cite{Lucero2020}.

\subsection{Steady-state condition}

We initialize in the standard Gibbs-Boltzmann equilibrium distribution,
\begin{align}
	P(\theta_{\rm o},\theta_{1},t=0) \propto \exp\left\{-\beta V(\theta_{\rm o},\theta_{1})\right\} \ .
\end{align}
We numerically integrate the 2D Fokker-Planck equation (see main text) with periodic boundary conditions using standard finite-difference methods~\cite{numerical_recipes}.
This evolves the joint probability distribution from a specified initial distribution to the steady-state distribution $P_{\mathrm{ss}}(\theta_{\rm o},\theta_{1})$.
Convergence to steady state is judged by the distribution remaining unchanged after evolution for $\Delta t = \num{e-3}$, as measured by the total variation distance:
\begin{align}
	\frac{1}{2}\iint &\dd{\theta_{\rm o}}\dd{\theta_{1}}\left|P_{\mathrm{ss}}(\theta_{\rm o},\theta_{1},t+\Delta t) - P_{\mathrm{ss}}(\theta_{\rm o},\theta_{1},t)\right|  \nonumber \\
	&< 10^{-16}.
\end{align}

\subsection{Setting the time scale}
We assign physical units by equating the simulation timescale and the physical timescale for analogous experiments.
We approximate ATP synthase as a sphere rotating around an axis through its center.
The rotational drag coefficient for a sphere of radius $r$ rotating in a fluid of viscosity $\eta$ is 
\begin{align}
	\zeta_r = 8 \pi \eta r^3 \ .
\end{align}
The viscosity of water is $10^{-9}$ pN s $\rm nm^{-2}$, and ATP synthase has a radius $\sim$15 nm.
The diffusion coefficient is found using the Einstein relation,
\begin{align}
	D_{\rm phys} = \frac{\kT}{\zeta_r},
\end{align}
where $k_{\rm B}$ is Boltzmann's constant, and $T = 300\, \rm K$ is room temperature.
The ratio of physical diffusion coefficient to simulation diffusion coefficient is
\begin{align}
	\frac{D_{\rm phys}}{D_{\rm sim}} = \frac{1.9 \cdot 10^6}{10^{-3}} \frac{\rm rad^2 s^{-1}}{\Delta \theta^2 \Delta t^{-1}} \ ,
\end{align}
for simulation grid spacing $\Delta \theta = \nicefrac{\pi}{180}\, \rm rad$ and simulation timescale $\Delta t$.
Setting this ratio to unity implies that the simulation timestep corresponds to
\begin{align}
	\Delta t = 6.7 \cdot 10^{-5}\, \rm s \ .
\end{align}

\section{Number of energy barriers}

\subsection{Varying \texorpdfstring{$n_{\rm o} = n_1$}{TEXT}}
\label{sec:barriers_n}
\begin{figure}[t]
	\centering
	\includegraphics[width=.5\textwidth]{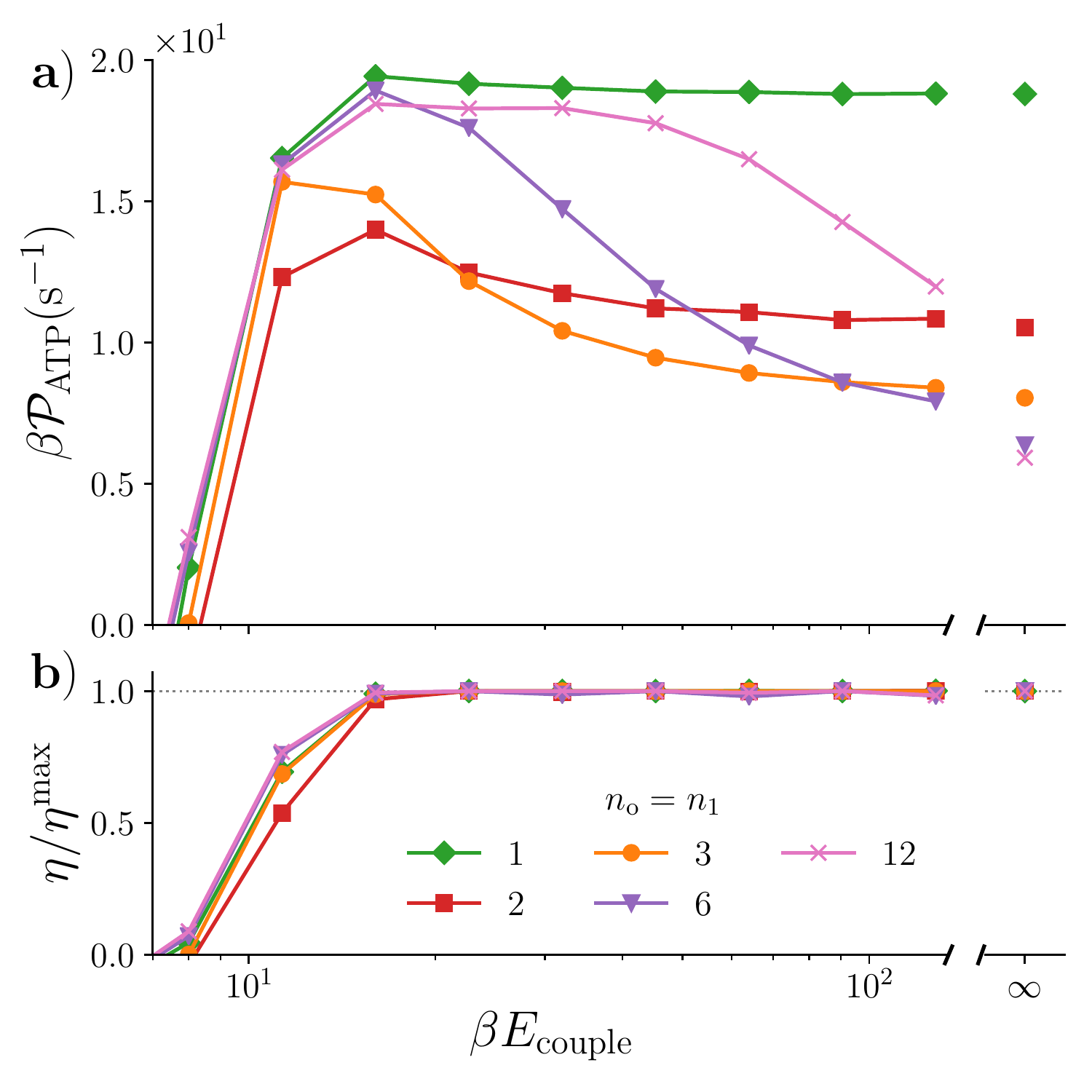}
	\caption{
		a) Output power as a function of coupling strength $\beta E_{\rm couple}$, for different numbers of barriers, with $n_{\rm o} = n_1$, no phase offset, barrier heights $\beta E_{\rm o} = \beta E_1 = 2$, and chemical driving forces $\mu_{\rm H^+} = 4\ \kT/ \rm rad$ and $\mu_{\rm ATP} = -2\ \kT/ \rm rad$. 
		b) Efficiency under the same conditions, scaled by the theoretical maximum efficiency $\eta^{\rm max} \equiv -\mu_{\rm ATP}/\mu_{H^+}$.
		Horizontal grey dotted line: maximum efficiency.
		Infinite coupling values are calculated using \eqref{eq:fluxrisken}.
	}
	\label{fig:power_eff_Ecouple_n}
\end{figure}

Here we vary the number of energy barriers, with the constraint $n_{\rm o} = n_1$.
Figure~\ref{fig:power_eff_Ecouple_n} shows output power and efficiency as a function of coupling strength for various numbers of barriers.
The output power curves in Fig.~\ref{fig:power_eff_Ecouple_n}a are similar to those in the main text. 
In particular, the curve of orange circles is identical to the case studied there.
Every curve in Fig.~\ref{fig:power_eff_Ecouple_n}a shows a maximum at some intermediate-strength coupling.
The peak is the most dramatic for six barriers, but quite subtle for a single barrier.
The efficiency, shown in Fig.~\ref{fig:power_eff_Ecouple_n}b, varies little with the number of barriers. 

Compared to the $n_{\rm o} = n_1 = 3$ result (orange circles in this plot), the output power for more barriers has a higher peak. 
The more barriers, the easier it is for one of the subsystems to diffuse ahead to a subsequent minimum since the subsystems can remain closer together, incurring a smaller `penalty' from the energetic coupling term. 
This also leads to wider peaks at these coupling strengths.
Moreover, when there is only a single minimum (and hence single barrier), it is counterproductive for \fo~to jump ahead, since it would end up in the same minimum again. 
This introduces slippage and reduced output power and efficiency.
It should be noted that a landscape with the prescribed barrier height, chemical driving force, and a single barrier actually does not have a local minimum.  
This leads to the green diamond curve being similarly shaped to the barrier-less case, which does not have a maximum at intermediate-strength coupling.

\begin{figure}[t]
	\centering
	\includegraphics[width=.5\textwidth]{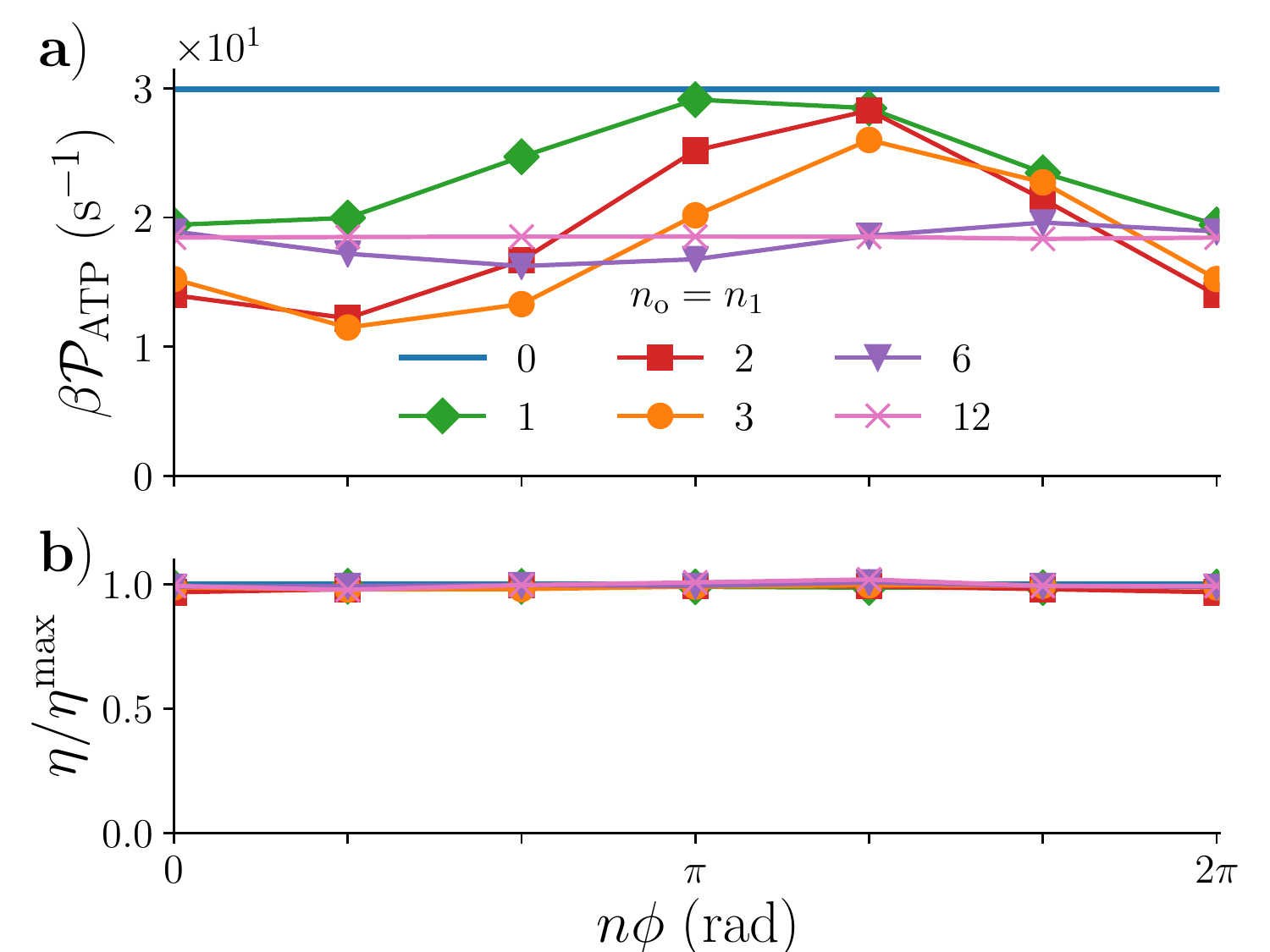}
	\caption{
		a) Output power as a function of scaled phase offset $n\phi$ between the subsystems with various numbers of barriers $n_{\rm o} = n_1$, at coupling strengths $\beta E_{\rm couple} = 16$, barrier heights $\beta E_{\rm o} = \beta E_1 = 2$, and chemical driving forces $\mu_{\rm H^+} = 4\ \kT/ \rm rad$ and $\mu_{\rm ATP} = -2\ \kT/ \rm rad$.
		b) Efficiency under the same conditions, scaled by the theoretical maximum efficiency.
	}
	\label{fig:power_eff_phi_n}
\end{figure}

Figure~\ref{fig:power_eff_phi_n} shows (a) output power and (b) efficiency as a function of the scaled phase offset $n\phi$, for various numbers of minima. 
The phase offset is scaled by a factor of $n_{\rm o} = n_1 = n$ to compare one period across all curves.
The output power varies as a function of the phase offset, though this variation is minimal for $n_{\rm o} = n_1 = 12$.
More barriers lead to smaller variation in output power.
More barriers means \fI~must overcome a smaller barrier height in a single hop to the next metastable state, since the tilt of the landscape stays the same.  
This leads to the effective barrier height decreasing, resulting in less variation in output power as the phase offset is varied.
At the same time, as more barriers are introduced the system slows down because there are more barriers to overcome.
The opposite is seen for fewer barriers: $n_{\rm o} = n_1 = 2$ has greater variation in output power, and the peak power is somewhat higher.
Effectively, a lower coupling strength is needed for a system with fewer barriers to approach the infinite-coupling power.

A single barrier, $n_{\rm o} = n_1 = 1$, leads to output power that varies slightly less with phase offset, but higher peak output power.
This is likely a consequence of the landscape not having any local minima at these parameters.

\subsection{Varying \texorpdfstring{$n_{\rm o}$}{TEXT}}
\begin{figure}[t]
	\centering
	\includegraphics[width=.5\textwidth]{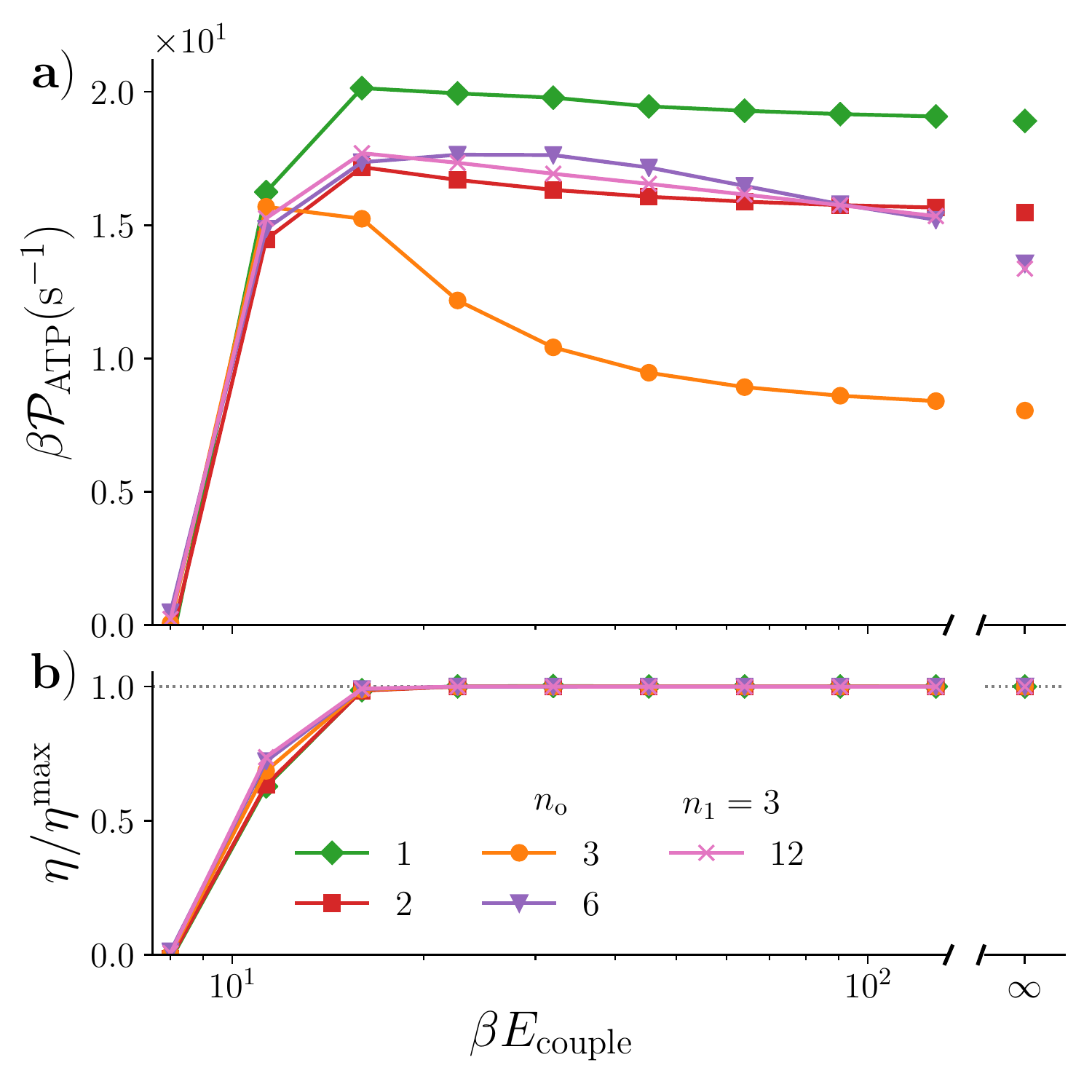}
	\caption{
		a) Output power and efficiency as a function of coupling strength $\beta E_{\rm couple}$ with various numbers of barriers $n_{\rm o}$, while keeping $n_1=3$ fixed.
		There is no phase offset, barrier heights are $\beta E_{\rm o} = \beta E_1 = 2$, and chemical driving forces are $\mu_{\rm H^+} = 4\ \kT/ \rm rad$ and $\mu_{\rm ATP} = -2\ \kT/ \rm rad$.
		b) Efficiency under the same conditions, scaled by the theoretical maximum efficiency. 
		Horizontal grey dotted line: maximum efficiency.
	}
	\label{fig:power_eff_Ecouple_n2}
\end{figure}

Figure~\ref{fig:power_eff_Ecouple_n2} shows output power as a function of coupling strength for a varying number of \fo~barriers, with 3 \fI~barriers.
For all $n_{\rm o}$, output power is maximized at intermediate-strength coupling.
The peak in output power is by far the most pronounced when $n_{\rm o}=3$ (orange circles), when the energy barriers in \fo~and \fI~align
and hence the trade-off is greatest between minimizing slip and loosening coupling sufficiently to capitalize on random fluctuations.
When $n_{\rm o} \neq n_1$, not all landscape barriers align with a landscape barrier of the other subsystem, leading to smaller effective barriers compared to the infinite-coupling limit.
Smaller peaks are easier to jump over, consequently they do not restrict the optimal coupling strength to the same degree.

\section{Barrier heights}
Once the barrier heights are sufficiently large to (in combination with tilts from driving forces) prevent any significant fraction of back steps, the output power is simply proportional to the rate of forward steps, and hence proportional to the exponential of the barrier height, $\exp[-\beta E^{\ddagger}]$. In this regime, the output power of any machine (regardless of driving forces and coupling strength) decreases with barrier height according to the same exponential decay. Thus once back steps are negligible, the ordering of machines by output power---and more specifically the coupling that optimizes output power---does not vary with barrier height. This physical intuition can be confirmed for tightly coupled subsystems, when the power is simply calculable by numerical integration (proportional to Eq.~\eqref{eq:fluxrisken}), permitting systematic exploration of its dependence on barrier height. 
	
\begin{figure}[t]
	\centering
	\includegraphics[width=.49\textwidth]{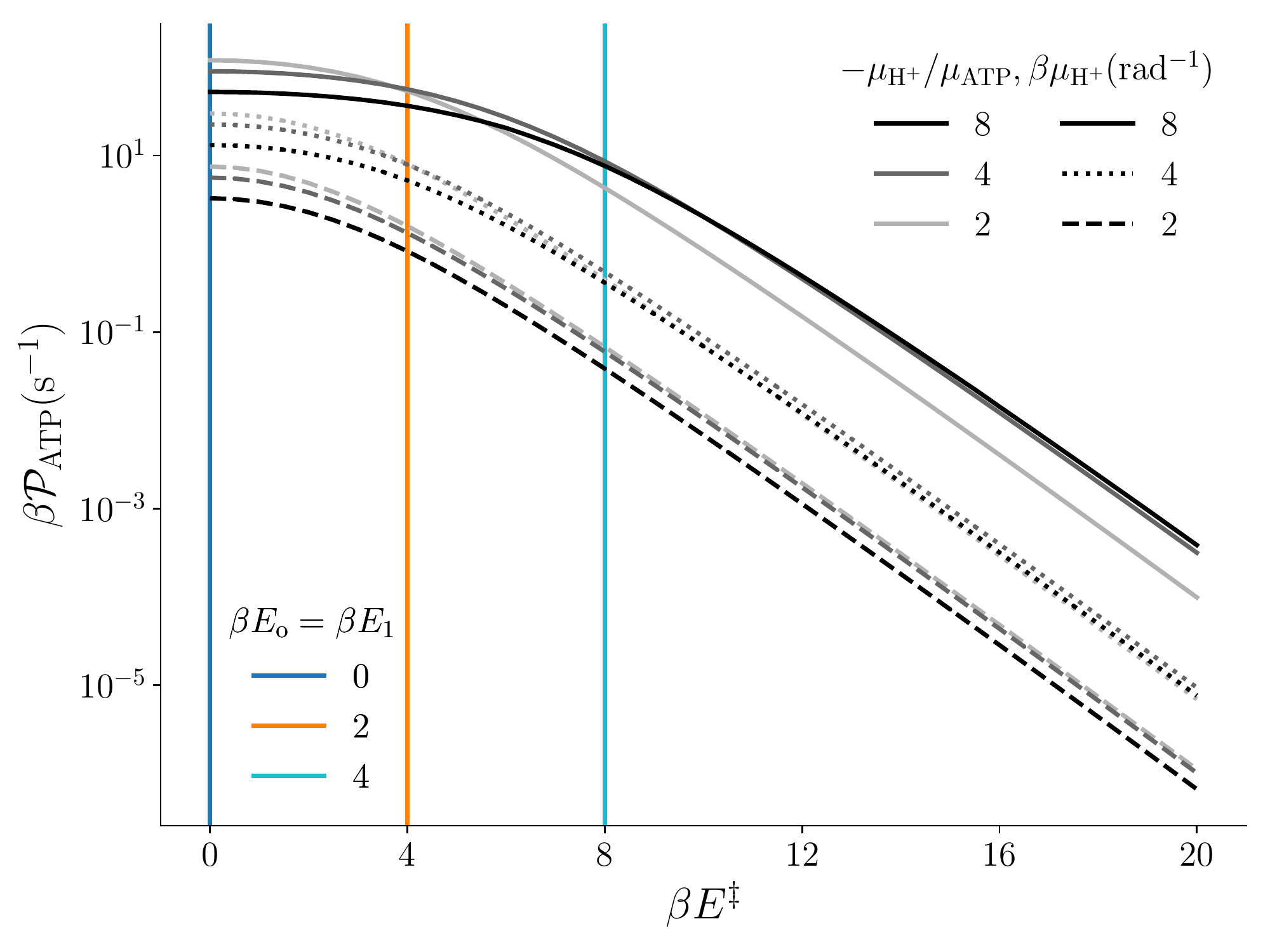}
	\caption{
		Output power as a function of barrier height $\beta E^\ddagger = \beta (E_{\rm o} + E_1)$ for the tightly coupled system.
		Different greyscale shades represent different ratios $-\mu_{{\rm H}^+}/\mu_{\rm ATP}$ of driving forces (corresponding to the columns in Fig.~3 of the main text), and different linestyles represent different proton driving forces $\beta \mu_{{\rm H}^+}$ (corresponding to the rows in Fig.~3 of the main text).
		Colored lines indicate barrier heights explored in the main text: $\beta E_{\rm o} = \beta E_1 = 0$ (dark blue), 2 (orange), or 4 (light blue).
	}
	\label{fig:Patp_barrier}
\end{figure}

Figure~\ref{fig:Patp_barrier} shows that beyond $\sim 8 \kT$ barriers, systems with all examined variations of driving forces (for no phase offset, $n=3$ states) have reached the regime of simple exponential dependence on barrier height.  Thus while the quantitative power-maximizing coupling may change somewhat with barrier height as it increases above 2 $\kT$ (already seen graphically in Fig.~3 
in the main text), the qualitative findings are likely robust to such variation.

\makeatletter
\providecommand{\doi}
{\begingroup\let\do\@makeother\dospecials
	\catcode`\{=1 \catcode`\}=2 \doi@aux}
\providecommand{\doi@aux}[1]{\endgroup\texttt{#1}}
\makeatother
\providecommand*\mcitethebibliography{\thebibliography}
\csname @ifundefined\endcsname{endmcitethebibliography}
{\let\endmcitethebibliography\endthebibliography}{}

\end{document}